\documentclass[prl,twocolumn,aps,showpacs,superscriptaddress]{revtex4}
\usepackage{graphics}
\usepackage{epsfig}
\usepackage{amsmath}
\usepackage{amssymb}
\usepackage{color}

\newcommand{\rom}[1]{\uppercase\expandafter{\romannumeral #1\relax}}

\begin{document}
\newcommand{\beq}{\begin{equation}}
\newcommand{\eeq}{\end{equation}}

\title{Non-Euclidean Origami}

\author{Scott Waitukaitis}
\affiliation{Huygens-Kamerlingh Onnes Lab, Leiden University, PObox 9504, 2300 RA Leiden, The Netherlands}\affiliation{AMOLF, Science Park 104, 1098 XG Amsterdam, The Netherlands}
\author{Peter Dieleman}
\affiliation{Huygens-Kamerlingh Onnes Lab, Leiden University, PObox 9504, 2300 RA Leiden, The Netherlands}\affiliation{AMOLF, Science Park 104, 1098 XG Amsterdam, The Netherlands}
\author{Martin van Hecke}
\affiliation{Huygens-Kamerlingh Onnes Lab, Leiden University, PObox 9504, 2300 RA Leiden, The Netherlands}\affiliation{AMOLF, Science Park 104, 1098 XG Amsterdam, The Netherlands}

\date{\today}

\begin{abstract} Traditional origami starts from flat surfaces, leading to crease patterns consisting of Euclidean vertices. However, Euclidean vertices are limited in their
folding motions, are degenerate, and suffer from misfolding. 
Here we show how non-Euclidean 4-vertices 
overcome these limitations by lifting this degeneracy,
and that when the elasticity of the hinges is taken into account,
non-Euclidean 4-vertices permit higher-order multistability.
%
%
We harness these advantages to design an origami inverter that does not suffer from misfolding and to physically realize a tristable vertex.
\end{abstract}


\maketitle

Origami provides a vast space to design novel mechanical metamaterials and folding devices \cite{Cheung:2014dm,Evans:2015et,Chen:2016bk, Schenk:2013kk, Wei:2013kn, Fang:2016bh, Waitukaitis:2015dk, Waitukaitis:2016by, Brunck:2016, Yasuda:2015eg, Silverberg:2014dn, Overvelde:2016gn, Overvelde:2017, Boatti:2017, Dudte:2016db, Filipov:2015,Miura:1985,Bertoldi:2017,Dieleman:2019}.  
The exceptional geometrical, shape-shifting and mechanical functionalities of these systems 
ultimate spring from the 
nonlinear folding motions of the building blocks of origami
\cite{Cheung:2014dm,Schenk:2013kk, Wei:2013kn, Silverberg:2014dn, Silverberg:2015gb, Waitukaitis:2016by, Yasuda:2015eg, Chen:2016bk, Overvelde:2016gn,BinLiu:2018iv,Overvelde:2017,Santangelo:2017dl,Dudte:2016db,Boatti:2017,Miura:1985,Bertoldi:2017,Dieleman:2019,Yasuda:2017,Howell:2016,Morgan:2016,Chen:2019,Brunck:2016,Chen:2018,Bende:2015, Waitukaitis:2018}.
These
building blocks are \textit{n}-vertices---units where \textit{n} straight folds connected to \textit{n} rigid plates meet at a point \cite{Schenk:2013kk, Wei:2013kn, Hanna:2014ed,Waitukaitis:2015dk,Waitukaitis:2016by, Fang:2016bh,Sareh:2015a,Sareh:2015b,Brunck:2016,Chen:2018,Miura:1985}. Most attention has been on Euclidean vertices---which in their unfolded state lie flat in the plane---and in particular 4-vertices as these have a single degree of freedom (Fig.~1a).
However, the folding motions of Euclidean 4-vertices are limited and degenerate. This degeneracy follows from fold-inversion symmetry: if the folded state  of a vertex, specified by the folding angles
$\{\rho_i\}$, constitutes a valid configuration, so does $\{-\rho_i\}$. Hence, the unfolded $\{\rho_i\!=\!0\}$ state of Euclidean 4-vertices is self-symmetric and non-generic. This leads to a dual branch structure, where two folding motions, $I$ and $II$, intersect at the flat state \cite{Waitukaitis:2015dk,Santangelo:2017dl,Stern:2017bya,Chen:2018}. In turn, this degeneracy makes Euclidean crease patterns prone to misfolding \cite{Stern:2017bya,Stern:2018,Tachi:2017}.

To lift this degeneracy we consider non-Euclidean 4-vertices, \textit{i.e.},~those where the sector angles, $\alpha_i$, add to $2\pi +\varepsilon$  (Fig.~1). It is known that for $\varepsilon \ne 0$, the folding branches of 4-vertices split and recombine into new branches \cite{Santangelo:2017dl}, and that 
non-Euclidean vertices can form `bowls' or `cones' in the case of a negative angular surplus ($\varepsilon<0$), and `saddles' in the case of a positive angular surplus ($\varepsilon> 0$) --- in contrast, Euclidean 4-vertices only admit `bird foot' mountain-valley (MV) patterns with one mountain and three valley folds (or vice versa, {\textit e.g.}~$\{\pm$$\mp$$\pm$$\pm\}$)\cite{Huffman:1976fw, Waitukaitis:2015dk, Waitukaitis:2016by, Abel:2016,footnoteMV}. However, a complete picture of the 
folding motions of non-Euclidean vertices is missing, and their potential has been overlooked.
  
Here we show how non-Euclidean vertices enhance the functionality of origami-based devices and materials.  First, we systematically evaluate the folding motions of non-Euclidean 4-vertices to show exactly how the branch splitting occurs, and find that non-Euclidean 4-vertices feature two distinct types of non-monotonic folding motions, in striking contrast to the monotonicity of the
folding motions of Euclidean 4-vertices  \cite{Huffman:1976fw,Waitukaitis:2015dk}.
We then consider how the absence of misfoldings leads to more robust nonlinear mechanisms, and leverage this to design an origami inverter. Finally, we show how branch splitting leads to a tuneable energy barrier which can be harnessed to obtain control the stability landscape, and physically realize a tristable vertex. Together, our work shows the versatility of non-Euclidean origami as building blocks for advanced mechanical metamaterials.

\begin{figure}[t!]
\includegraphics[]{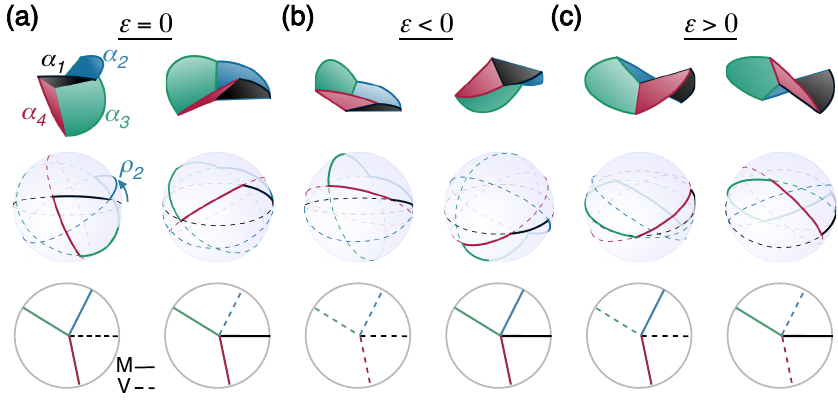}
\caption{A 4-vertex is specified by  four sector angles, $\alpha_i$, and its folded states are described by four folding angles, $\rho_i$, which are the complements to the dihedral angles between plates $i$ and $i\!-\!1$. Mountain and valley folds correspond to $\rho_i>0$ and $\rho_i<0$ respectively. (a) Folded states of a Euclidean 4-vertex (top), representation on the Euclidean sphere (middle), and corresponding MV patterns (bottom).   (b) Non-Euclidean vertex with a negative angular surplus in a `bowl' or `cone'  configuration.  (c) Non-Euclidean vertex with a positive angular surplus in `saddle' configurations.}
 \label{fig:geometry}
\end{figure}

\begin{figure*}[t!]
\includegraphics[]{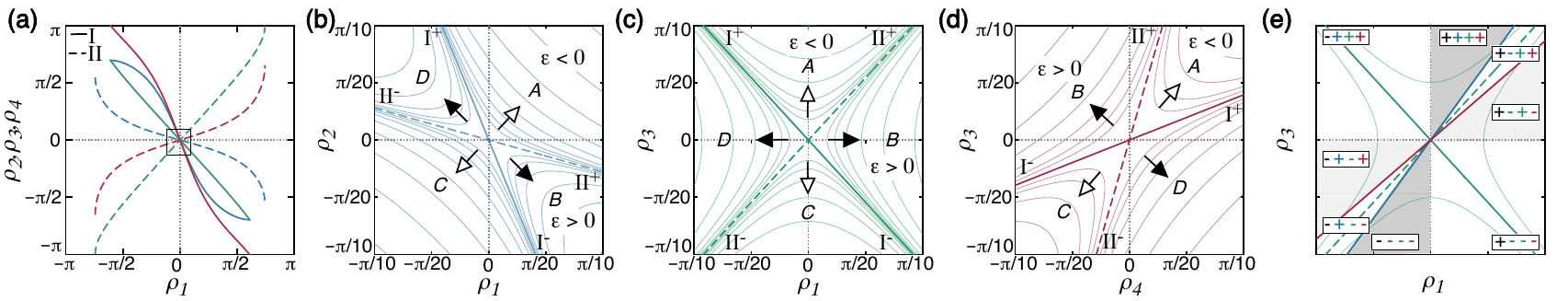}
\caption{(a) Folding curves $\rho_2$ (blue), $\rho_3$ (green), and $\rho_4$ (magenta) vs.~$\rho_1$ for branch I (solid) and II (dashed) of a Euclidean vertex with sector angles  $\alpha_i = \{ \pi/3, \pi/2, 3\pi/4, 5\pi/12\}$.  The box in the center highlights zoomed views corresponding to panels (b-e), which also include curves for non-Euclidean vertices created by uniform shrinkage/expansion of $\alpha_i$. Open and closed arrows indicate splitting directions for $\varepsilon<0$ and $\varepsilon>0$, respectively.  The branch endpoints I$^-$, I$^+$, II$^-$, II$^+$ are indicated, as well as the resulting non-Euclidean branch designations $A$, $B$, $C$ and $D$ (see Table \ref{tafeltje}).  (e) Schematic plot of $\rho_3$ vs.~$\rho_1$ with different MV assignments indicated with shading allows to visualize the sequence of MV patterns along each branch.}
\label{fig:mercator_b}
\end{figure*}

\begin{figure}[t]
\includegraphics[]{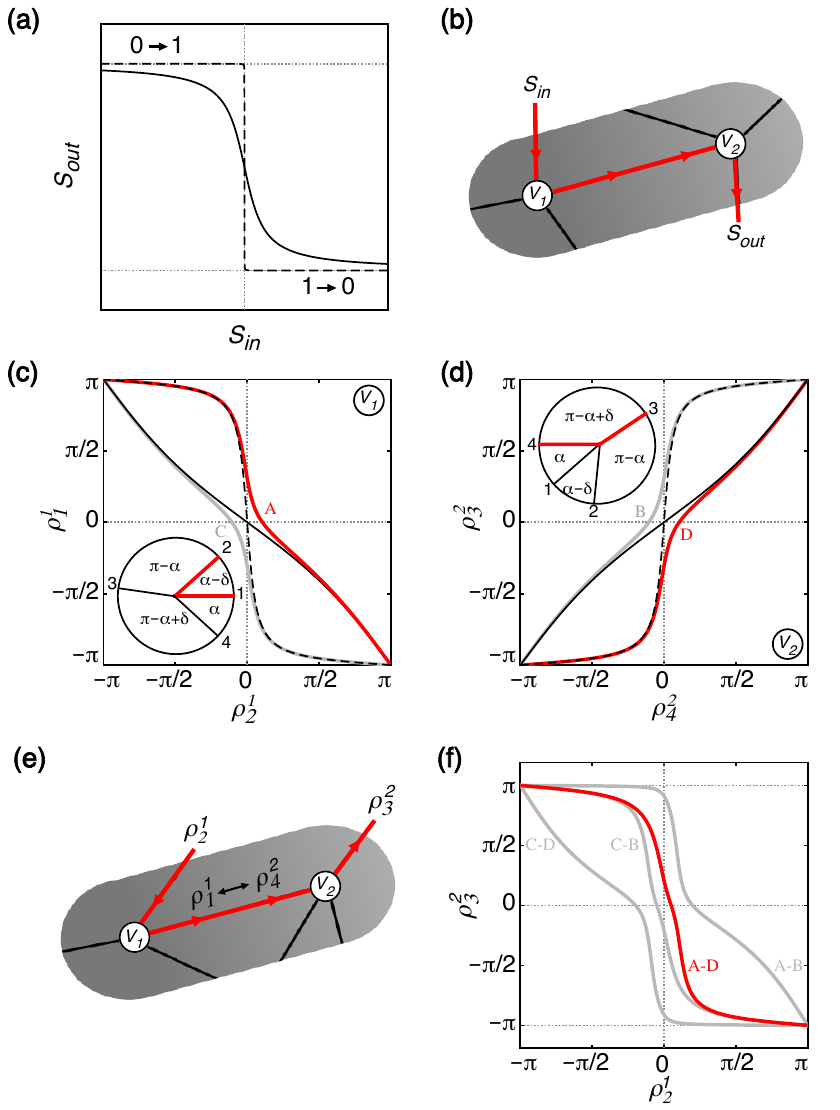}
\caption{(a)  An inverter maps a high input signal, $S_{in}$$\sim$$1$, to a low output signal, $S_{out}$$\sim$$0$, and vice versa.  (b)  Composite design of two non-Euclidean 4-vertices, $V_1$ and $V_2$, with  input/output  corresponding to fold angles $S_{in}$ and $S_{out}$ (red).  (c)  Curve for vertex $V_1$ on branch $A$ (red); folding curves for the corresponding Euclidean vertex (black solid/dashed lines) and other branch of non-Euclidean vertex (gray) are also shown.  (d)  Curves for vertex $V_2$.  (e)  Origami inverter design.  (f)  Composite inverter curve (red) and `spurious' fold curves when one or both vertices are forced on their other branch (grey). Parameters for the target curve are $|\varepsilon|=\pi/50$, $\alpha=\pi/4$ and $\delta = 0.1$.}
 \label{fig:inverter}
\end{figure}

{\em Branch splitting and folding curves.}---We start by determining the qualitative nature of the folding branches for generic vertices with $\varepsilon \ne 0$. 
For Euclidean vertices, the unique folds (the one with opposite folding angle from the others) follow from inequalities on the sector angles $\alpha_i$ \cite{Waitukaitis:2015dk,Waitukaitis:2016by}, and we orient these vertices such that  $\alpha_2>\alpha_4$ and folds 1 and 2 are the unique folds on branch $I$ and $II$, respectively (See S.I. for details).  
The non-Euclidean vertices we consider are nearly flat (small $|\varepsilon|$) and are derived from shrunken or expanded Euclidean ones.
Maintaining our conventions during shrinkage/expansion, we isolate generic features even while working with specific examples.

We start by examining the branch splitting for a family of vertices with sector angles $ \alpha_i$$=$$(1$$+$$\varepsilon/(2 \pi)) \{ \pi/3, \pi/2, 3\pi/4, 5\pi/12\}$.  We numerically calculate the folding curves, $\rho_i(\rho_j)$, for $\varepsilon=0$ and several values of $\varepsilon\ne0$ and plot these in Fig.~\ref{fig:mercator_b} (for derivations of the folding curves see, \textit{e.g.}, \cite{Evans:2015et,Waitukaitis:2015dk}).  For $\varepsilon \ne 0$, distortion causes the Euclidean branches $I$ and $II$ to split into four disconnected branches that we label $A$-$D$ [Figs. \ref{fig:mercator_b}(b-d)].  For $\epsilon <  0$ the vertex is either on branch $A$ or $C$, and for $\epsilon >  0$ the vertex is on branch $B$ or $D$ (see S.I.). Pairs of branches are related by the fold-inversion symmetry discussed above. Clearly, the folding motions of non-Euclidean 4-vertices do not have any branch points and are smooth.  Assuming rigid folding, a non-Euclidean vertex's branch designation is fixed and cannot change (\textit{e.g.}, a rigid vertex on branch $A$ cannot switch to branch $C$).  As a consequence, a non-Euclidean vertex is specified by both its sector angles and its branch label.

Although corresponding to a specific choice of sector angles, the qualitative features of this splitting (up/down, right/left, four branches $A$-$D$) are completely general. To show this, and to
connect the branches to qualitative shapes---cones, bowls, saddles, and bird's feet---we consider the sign of the folding angles along each branch, starting from their maximally folded states.
We label the endpoints of the Euclidean folding
branches as $I^+$:$\{-+++\}$, $I^-$:$\{+---\}$, $II^+$:$\{+-++\}$ and $II^-$:$\{-+--\}$ [Fig.~\ref{fig:mercator_b}(b-d)]. For small $\varepsilon$, the non-Euclidean folding branches must have the endpoints close to their Euclidean parent, with the same MV patterns.
The six possible folding branches of 4-vertices therefore follow from connecting pairs of endpoints. Euclidean folding branches $I$ and $II$ connect $I^+ \leftrightarrow I^-$ and $II^+ \leftrightarrow II^-$, respectively, and intersect at the flat state.  The remaining four endpoint-pair combinations correspond to the non-Euclidean folding branches $A$-$D$ that avoid the flat state (Fig.~\ref{fig:mercator_b}). We determine the MV patterns on each branch by noting that generically, one cannot have two fold angles pass through zero, or in other words, folds change between mountain and valley one-by-one along these branches.  For example, on branch $A$, the endpoints $\{-+++\}$ and $\{+-++\}$ must be connected by the `bowl' $\{++++\}$, and therefore this branch must correspond to $\varepsilon <0$; similarly, on branch $B$, the endpoints $\{+-++\}$ and $\{+---\}$ must be connected by the `saddle' $\{+-+-\}$ and this branch must have $\varepsilon >0$; we summarize all six branches in Table 1 \cite{footnotetable}.  (Patterns with cyclic permutations of $\{++--\}$ are not allowed as they create intersections---see S.I.)

\begin{table}[b]
\begin{tabular}{l|l|l|l}
& Endpoints  & MV patterns  &  \\\hline
$I$ & $I^+ \! \leftrightarrow\! I^-$ &  $\{-+++\} \! \leftrightarrow \! \{0,0,0,0\} \! \leftrightarrow \! \{+---\}$ & $\varepsilon \!=\!0$\\
$II$ & $II^+ \! \leftrightarrow \! II^-$ & $ \{+-++\} \! \leftrightarrow \! \{0,0,0,0\} \! \leftrightarrow \! \{-+--\}$ & $\varepsilon \!=\!0$\\
$A$ & $I^+ \! \leftrightarrow \! II^+$ & $ \{-+++\} \! \leftrightarrow \! \{++++\} \! \leftrightarrow \! \{+-++\} $ & $\varepsilon \!<\!0$ \\
$B$ & $II^+ \! \leftrightarrow \! I^-$ & $ \{+-++\} \! \leftrightarrow \! \{+-+-\} \! \leftrightarrow \! \{+---\}$ &
$\varepsilon \!>\!0$ \\
$C$ & $I^- \! \leftrightarrow \! II^-$ & $ \{+---\} \! \leftrightarrow \! \{----\} \! \leftrightarrow \! \{-+--\}$ &
$\varepsilon\! <\!0$ \\
$D$ & $II^- \! \leftrightarrow \! I^+$ & $ \{-+--\} \! \leftrightarrow \! \{-+-+\} \! \leftrightarrow \! \{-+++\}$ &
$\varepsilon \!>\!0$ \\
\end{tabular}
\caption{{Folding branches and MV patterns}.}
\label{tafeltje}
\end{table}

Now that we have established how branches $A-D$ arise, we can determine how the folding motions along these branches differ qualitatively from those of Euclidean 4-vertices.
For a Euclidean vertex, the folding relations between any pair of $\rho_i$ on branches $I$ or $II$ are always monotonic and always capable of having positive or negative sign  \cite{Waitukaitis:2015dk}.  
However, Table 1 and Fig.~2 show that for non-Euclidean vertices the qualitative nature of the folding motion depends on the branch and the pair of folding angles considered.
First, monotonic folding curves with negative slope occur between the unique folds 1 and 2 for $\varepsilon <0$, or with positive slope between the non-unique folds 3 and 4 for $\varepsilon > 0$ [Fig.~\ref{fig:mercator_b}(b,d)]. Between a unique (1,2) and a non-unique (3,4) fold, all curves are non-monotonic.  For $\varepsilon<0$, the unique fold monotonically changes sign, whereas the non-unique fold is non-monotonic and has a fixed sign; for $\varepsilon>0$, the non-unique fold monotonically changes sign and the unique fold is non-monotonic with fixed sign (see Fig.~\ref{fig:mercator_b}(c) for one such example). Finally,  between folds 1 and 2 for $\varepsilon >0$, and between the folds 3 and 4 for $\varepsilon < 0$, both folds have a fixed sign and are non-monotonic [Fig.~\ref{fig:mercator_b}(b,d)].
These qualitatively different folding motions open up new design possibilities for folding mechanisms, as
well as ratioanl design of multistable structures, as we show below.

{\em Designer Mechanisms.}---We now show how the folding curves of a non-Euclidean vertex can be used to design nonlinear mechanisms.  We illustrate this general point by designing an inverter, where a high input signal  is mapped to a low output signal and vice versa [Fig.~\ref{fig:inverter}(a)]. In our origami inverter, the input and output signals correspond to fold angles.  A Euclidean Miura vertex with sector angles $\alpha_i = \{ \alpha, \pi-\alpha, \pi-\alpha, \alpha \} $ has one branch with a curve ($\rho_i$ {\textit vs.}~$\rho_j$) that reproduces a step function with infinite slope and sharp corners  \cite{Waitukaitis:2016by,Miura:1985,Sareh:2015a,Sareh:2015b,Wei:2013kn,Papa:2008}---close to the behavior desired for an inverter.  However, the intersection of this branch at the flat state with the other `distractor' branch precludes the necessary one-to-one functionality.

To resolve this, we consider near-Miura, non-Euclidean candidates given by the sector angles $\alpha_i = (1 \pm |\varepsilon |/(2 \pi)) \{ \alpha-\delta, \pi-\alpha, \pi-\alpha+\delta, \alpha \} $.  The parameter $\delta$ breaks the `Miura'-symmetry, allowing us to (i) stay within our generic framework and (ii) control the sharpness of the step-function. The general qualitative properties of 
the folding curves of non-Euclidean vertices ({\textit e.g.}~the curvature of the monotonic branch)
prevent a single non-Euclidean vertex from achieving a folding branch with an `S'-shape and
inverter functionality. However, the absence of misfolding allows us connect multiple vertices without possible branch switching.  By
joining two vertices ($V_1$ and $V_2$) and choosing their design and branches appropriately, we
can achieve an origami inverter [Fig.~\ref{fig:inverter}(b)] as follows. Considering the input angle ($\rho_{in}^1$), the connecting angle ($\rho_{out}^1$=$\rho_{in}^2$) and the output angle ($\rho_{out}^2$), the transfer curve is given by
\begin{equation}
F\big{(}\rho_{in}^1 \big{)} = \rho_{out}^2 \Big{(} \rho_{out}^1 \big{(} \rho_{in}^1 \big{)} \Big{)}~.
\end{equation}
The slope of a inverter curve must be negative, which accounting for the previously discussed (non-)monotonic nature of the non-Euclidean branches, is possible if for one vertex we use folds 1 and 2 and take $\epsilon > 0$, and for the other we use folds 3 and 4 and take $\epsilon <0$.  We therefore use as our input signal $\rho_2^1$ of vertex $V_1$ with $\alpha_i = (1 - |\varepsilon |/(2 \pi)) \{ \alpha-\delta, \pi-\alpha, \pi-\alpha+\delta, \alpha \} $, and as our output signal $\rho_3^2$ of vertex $V_2$ with $\alpha_i = (1 + |\varepsilon |/(2 \pi)) \{ \alpha-\delta, \pi-\alpha, \pi-\alpha+\delta, \alpha \} $ (connected by folds $\rho_1^1$=$\rho_4^2$).  This yields the composite structure shown in Fig.~\ref{fig:inverter}(e), which produces a folding curve closely matching the target [Fig.~\ref{fig:inverter}(f); parameters in caption].   Crucially, the other compound folding curves, obtained by changing the branch of $V_1$, $V_2$, or both, are well separated, even for small (1\%) deviation of flatness. This illustrates the potential of non-Euclidean origami for well-defined designer mechanisms that circumvent the problem of distractor branches.

{\em Energy Landscapes of non-Euclidean 4-vertices.---} 
The branch splitting has significant consequences for the energy and stability landscapes of non-Euclidean 4-vertices. Modeling the hinge elasticity with torsional springs and assuming the plates are {\em rigid}, the energy of a vertex is given by \cite{Wei:2013kn,Silverberg:2015gb,Waitukaitis:2015dk,Brunck:2016,Silverberg:2014dn,Joules:2019,Brunck:2016,Lechenault:2014,Liu:2017,Filipov:2017}:
\begin{equation}
E=\frac{1}{2}\sum_{i=1}^4\kappa_i(\rho_i-\bar{\rho}_i)^2~.
\end{equation}
Here $\kappa_i$ are the torsional spring constants and $\bar{\rho}_i$ the rest angles of each fold.  For a Euclidean vertex, the existence of two folding branches means that there are also two energy curves.  These intersect at the flat state and, as we showed previously, this has the implication that generic Euclidean vertices are at least bistable.  
While more minima are possible---theoretically up to six \cite{Waitukaitis:2015dk} if there is sufficient freedom in the spring parameters---these populate a vanishingly small volume of design space and are too shallow to permit physical implementation \cite{Waitukaitis:2015dk}.

\begin{figure}[t]
\includegraphics[]{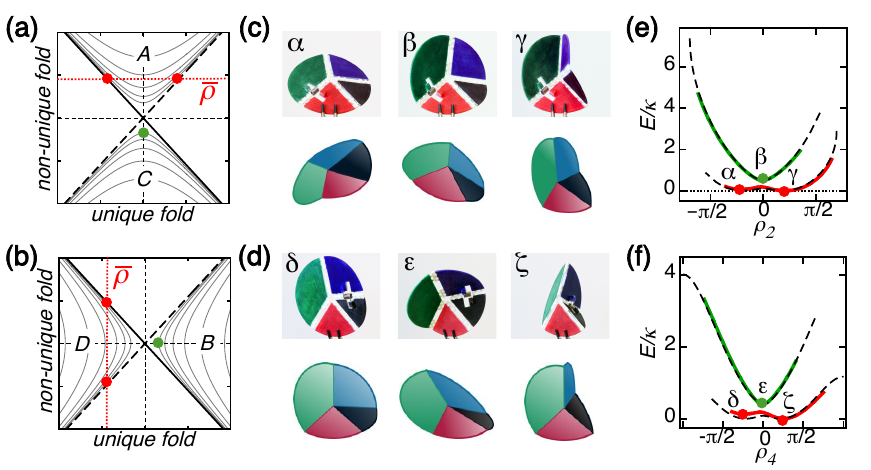}
\caption{(a) For a vertex with $\varepsilon <0$, placing a single spring on a non-unique fold with rest angle $\bar{\rho}>0$ (red line) yields two $E=0$ minima on the $A$-branch (red dots) and one $E\ne 0$ `frustrated' minimum on the $C$-branch (green dot).  (b) Similar for a vertex with $\varepsilon > 0$ and a spring placed on a unique fold. (c) Physical samples (top row) and numerical model (bottom row) for a 4-vertex $\alpha_i= 119/120 ~\{ \pi/3, \pi/2, 3\pi/4, 5\pi/12 \}$ augmented by a torsional spring on fold 4, in each of its three stable states labeled $\alpha-\gamma$. Plates are colored by the convention set in Fig.~1. (d) Physical samples (top row) and numerical model (bottom row) for a 4-vertex $\alpha_i=121/120~ \{ \pi/3, \pi/2, 3\pi/4, 5\pi/12 \}$ augmented by a torsional spring on fold 2, in each of its three stable states labeled $\delta-\zeta$. (e-f) By 3D-printing vertices and adding springs, we are able to verify the effectiveness of these strategies.  Panel (e) compares the theoretical (dashed lines) and experimentally measured (green and red lines) curves for the negative-surplus vertex.  Panel (f) shows corresponding curves for a positive-surplus vertex. Both panels also indicate the minima $\alpha-\zeta$.}
 \label{fig:energyScheme}
\end{figure}

To understand how non-Euclidean vertices differ, we consider what happens when a single torsional spring is placed on one of the folds.  A Euclidean vertex will be able to reach a zero energy minima on both branches $I$ and $II$, leading to two stable states.  For a nearby non-Euclidean vertex, a different scenario emerges, and the nature of the folding motions suggests that the placement of the spring is critical.  If $\epsilon<0$ and the spring is on a unique fold, then regardless of the branch the vertex is on ($A$ or $C$) only one stable (zero-energy) minima is accessible.  (A similar case holds for $\epsilon>0$ with a spring on a non-unique fold.) However, if the spring is placed on a non-unique and $\varepsilon<0$, as in Fig.~\ref{fig:energyScheme}(a), then one branch ($A$ if $\bar{\rho}>0$) will have two zero-energy minima, while the other will have a frustrated, finite energy minima near to the flat state.  (Again, a similar situation happens for branches $B$, $D$ when $\epsilon>$0 and the spring is on a unique fold.) So long as the corresponding energy barrier can be exceeded, \textit{i.e.}~the plates are not \textit{too} rigid and can be bent or stretched, the vertex can `pop through' the flat state \cite{Silverberg:2014dn}.  This leads to a simple rule for creating robust {\em tristable} vertices:  (1) for $\epsilon<0$, place a single spring on a non-unique fold; (2) for $\epsilon>0$, place a single spring on a unique fold.

For a physical realization, we 3D print non-Euclidean vertices with sector angles $\alpha_i = (1+\varepsilon/(2\pi) )\{ \pi/3, \pi/2, 3\pi/4, 5\pi/12\}$ out of ABS plastic, which allows for a small amount of elastic deformation (for fabrication details, see S.I.).  With an appropriate value of angular surplus ($\varepsilon \approx \pm 0.0083$) these vertices exhibit robust pop through behavior.  When additionally paired with a torsional spring---which can be mounted on the vertex in 3D printed holes---one of the two branches can be made bistable, whereas the other branch is monostable with a  frustrated minima---validating the strategy for tristable vertices  Fig.~\ref{fig:energyScheme}(c-d)].  For both $\varepsilon>0$ and $\varepsilon<0$, we have measured the elastic energies along both branches, and find that it compares well to the theoretical prediction based on the geometric design and torsional stiffness  of our spring [Fig.~\ref{fig:energyScheme}(e-f)]. For full details on how we fabricate the vertices and measure the energy curves, we refer the reader to the S.I.  Hence, for non-Euclidean vertices with finite hinge and plate elasticity---the situation most relevant to many applications---independently tuning the energy barrier between branches and the energy landscape on the branches results in a novel strategy for multi-stable origami.

{\em Conclusion and Discussion.---}
Euclidean 4-vertices sit at a critical plane in parameter space, and undergo a bifurcation when they are made non-Euclidean by shrinking or extending their sector angles. The associated branch splitting lifts the degeneracy in the folding motions, leads to novel mountain-valley patterns, and yields novel, non-monotonic folding curves. Non-Euclidean 4-vertices do not suffer from distractor branches, and we have shown how to use this to design a nonlinear mechanism. Non-rigid, non-Euclidean 4-vertices can exhibit a pop through between branches, offering a simple pathway to tristable structures.  While we have focussed on single and dual non-Euclidean 4-vertices, we point out that a recent design methodology, initially developed for flat 4-vertices, can readily be adapted to design a wide variety of periodic and spatially textured crease patterns that combine non-Euclidean 4-vertices with positive and negative angular surplus \cite{Dieleman:2019}. One interesting question for the future is  how the non-monotonic folding motions of individual 4-vertices affects those of larger folding patterns. A second question is  how to extend our results to higher-$n$ vertices, and in particular whether we can use higher-$n$ vertices to design more complex mechanisms \cite{Song:2019}.

{\em Acknowledgements}  We thank C.~Coulais, C.~Santangelo, I.~Cohen and A.~Evans for productive discussions, and M.~Mertens for exploratory studies on the origami inverter.  We acknowledge funding from the Netherlands Organization for Scientific Research through grants VICI No.~NWO-680-47-609 (M.v.H.~and S.W.) and VENI No.~NWO-680-47-453 (S.W.).

\setcounter{figure}{0}
\renewcommand{\figurename}{SUPPL.~FIG.}
\renewcommand\thefigure{\arabic{figure}}
\setcounter{table}{1}
\renewcommand{\tablename}{SUPPL.~TABLE}
\renewcommand\thetable{\arabic{table}}

\begin{center}
{\huge{ \textbf Supplemental Information}}
\end{center}

\section{Conventions} 

We consider `generic' vertices where the sector angles meet three conditions:  (i) no two angles are equal, (ii) no pairs of adjacent angles add to $\pi$, and (iii) all angles are less than $\pi$. 
We maintain a counter-clockwise orientation of the vertex and defining the fold angles with respect to the right hand rule; positive/negative folding angles correspond to `valleys/mountains'.  We routinely denote MV patterns by the four signs of $\rho_i$ (\textit{e.g.~}$\{+,-,-,-\}$), keeping in mind that all vertices exhibit fold-inversion symmetry---if $\{\rho_i\}$ is a valid fold configuration, so is $\{-\rho_i\}$. The two branches of folding motion of generic Euclidean 4-vertices each correspond to a MV pattern with three folds of the same sign and one `unique' fold with the opposing sign (Fig.~1a). Which two folds can be unique follows from inequalities on the sectorangles $\alpha_i$ \cite{Waitukaitis:2015dk}. We orient Euclidean vertices such that folds 1 and 2 are the unique folds and such that $\alpha_2>\alpha_4$. When working with non-Euclidean vertices derived from shrunken or expanded Euclidean ones, we maintain this designation for the folds. These conventions ensure that the generic features of branches and folding curves are preserved for generic vertices.  Specifically, the slopes $d\rho_j / d\rho_i$ are maintained, as well as the structure of the branches $I$ and $II$ (and subsequently $A$, $B$, $C$, and $D$ for non-Euclidean vertices).

\section{Geometry of non-Euclidean 4-vertices}

\begin{figure*}[t!]
\includegraphics[width = \textwidth]{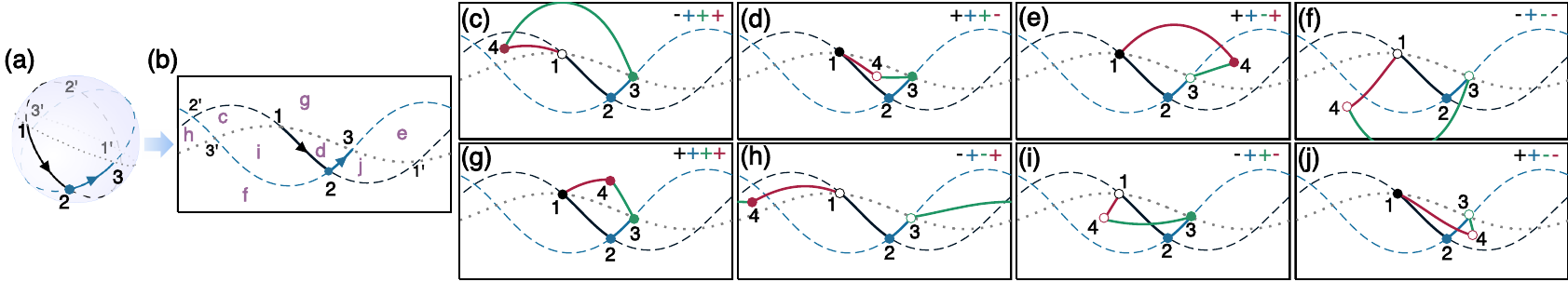}
\caption{(a) Random placement of the points $x_1$, $x_2$ and $x_3$, representing the folds 1, 2, and 3, on the Euclidean sphere.  Also indicated are the plate edges (solid lines), antinodal points $x'_1$, $x'_2$, and $x'_3$, and great arcs passing through these points (dashed and dotted lines). (b) Same as (a) but on a Mercator mapping, showing how the three great circles partition the sphere in eight spherical triangles that we label $c$- $j$. (c-j)  A definite MV-pattern emerges when $x_4$ is placed in any one of the eight triangles (fold angle sign represented by closed/open dots).  }
\label{fig:mercator_t}
\end{figure*}

{\em Mountain-Valley arrangements.---}  We now determine which MV patterns can arise in non-Euclidean 4-vertices, both for $\varepsilon<0$ and $\varepsilon >0$.  To systematically map out all possibilities, we represent generic 4-vertices (Euclidean or non-Euclidean) as spherical linkages, mapping each fold $i$ to a point $x_i$ on the Euclidean sphere $S$ and connecting these points by geodesic arcs of length $\alpha_i$ \cite{Huffman:1976fw}.  We then randomly place the four points $x_i$ on the sphere [Fig.~\ref{fig:mercator_t}(a)].  For clarity, we represent the sphere $S$ on flat paper with the Mercator mapping [Fig.~\ref{fig:mercator_t}(b)]. Without loss of generality we start with $x_1$, $x_2$ and $x_3$ and their antinodal points $x'_i$.  For definiteness, we first assume that $\rho_2 >0$---the case $\rho_2 <0$ is related by fold inversion symmetry. We focus on sector angles $\alpha_i \le \pi$, connect $x_1$ and $x_2$ by their shortest geodesic arc, $\overline{x_1x_2}$, and indicate the great circle coincident with this arc, $C_{12}$. We then repeat this procedure for the pairs $x_1/x_3$ and  $x_2/x_3$.  The great circles $C_{12}$, $C_{13}$ and $C_{23}$ partition $S$ in 8 triangular sectors labeled $(c-j)$ (Fig.~\ref{fig:mercator_t}b). Irrespective of our choice of $x_1$, $x_2$ and $x_3$, a definite MV pattern emerges when $x_4$ is placed in any of these sectors, as the great circles determine the sign of $\rho_i$:  $\rho_1>0$ when $x_4 \in \{d, e, g,  j\}$ and $\rho_1<0$ otherwise; $\rho_3>0$ when $x_4 \in \{c, d, g,  i\}$;  $\rho_4>0$ when $x_4 \in \{c,e,g,h\}$. Taken together, and reintroducing fold-inversion symmetry by relaxing the condition $\rho_2>0$, we find that the full list of possible mountain valley assignments, accounting for relabeling  symmetries (\textit{i.e.}, $\rho_i \leftrightarrow \rho_{i+1}$ etc), is as follows.  In sectors (c,d,e,f), we find `bird foot' configurations ($\{\pm,\mp,\mp,\mp\}$); in sector (g) we find `bowls' and `cones' ($\{\pm,\pm,\pm,\pm\}$), and in sector (h) we find `saddles' ($\{\pm,\mp,\pm,\mp\}$). Finally, the vertices obtained by placing $x_4$ in sectors (i,j) correspond to ($\{\pm,\pm,\mp,\mp\}$) but necessarily involve self-intersections, that can only be circumvented when we allow  $\alpha_i > \pi$.

The location of $x_4$ restricts the value of $\varepsilon$ in some sectors.  First, we note that the angular surplus has no fixed sign in cases (c-f), which implies that non-Euclidean vertices can  have the standard `Huffman' motif, as expected from continuity. However, one can show that  in  sector (g), $\varepsilon <0$, whereas in sector (h), $\varepsilon >0 $. To determine the sign of the angular surplus for vertices in configuration (g) and (h), we proceed as follows. First, given a spherical triangle $ABC$ and a point $D$ within this triangle, we note that $AD + DC < AB + BC$. For case (g), the relevant triangle is span by $x_1, x_3$ and $x'_2$, and so it follows that $x_1 x_4 + x_4 x_3 < x_1 x'_2 + x'_2 x_3$. Hence, it follows that the sum of the sector angles, $x_1 x_2 + x_2 x_3 + x_3 x_4 + x_4 x_1  < x_1 x_2 + x_2 x_3 + x_1 x'_2 + x'_2 x_3$; by definition, $x_1 x_2 + x_1 x'_2 = \pi $ (and permutations thereof), so that we find for case (g) that $x_1 x_2 + x_2 x_3 + x_3 x_4 + x_4 x_1  < 2 \pi$, and hence $\varepsilon <0$.

For case (h), we note that $x_1 x_4 > x_1 x'_2$ (otherwise point 4 would have been in triangle (c), (i), (d) or (g)), and similarly, $x_3 x_4 > x_3 x'_2$. Hence  $x_1 x_2 + x_2 x_3 + x_3 x_4 + x_4 x_1  >  x_1 x_2 + x_2 x_3 + x_3 x'_2 + x_1 x'_2 = 2\pi $, so that in case (h), $\varepsilon > 0$.

Finally, we note that case (d) also may appear to have $\varepsilon <0$, but this is not necessarily the case when $x_1 x_2$ and $x_2 x_3$ become large; indeed applying the same reasoning as above, we find that $x_1 x_2 + x_2 x_3 + x_3 x_4 + x_4 x_1  < 2 \left[ x_1x_2+ x_2x_3 \right]$ which does not restrict the sign of $\varepsilon$.

\section{Tristable vertex experiments}

\subsection{Fabrication}  Tristable vertices were 3D printed with a Stratasys Fortus 250 MC, which is capable of printing ABS plastic, as well as a sacrificial ABS-like plastic, with a layer thickness of 0.18 mm and an XY-resolution of better than 0.24 mm.  The sacrificial material serves as a scaffold and allows us to print non-flat vertices.  The scaffold is dissolved after printing by putting the structure in a 70$^\circ$ C NaOH solution.  With this technique, we are able to print non-flat vertices with a large range of angular surplus $\varepsilon$.

Our vertices are 150 mm in diameter, consisting of four plates which are 3.0 mm thick.  Each adjacent pair of plates is connected by hinges directly printed in a joined configuration.  The hinges consist of two conical holes attached to one plate, and two opposing conical pins attached to the opposing plate.  With just enough clearance to allow for rotation once extracted from the support material, no post-assembly is required.  The axes of rotation of all these hinges meet at the center of the vertex.  The main experimental limitation is the finite maximal folding angle ($\sim$2.65 rad) due to the formation of self-contacts between plates for high folding angles.

Each adjacent plate pair is designed to accommodate a torsional spring in 1.14 mm diameter holes on their sides near the hinge, which can be inserted after printing.  The holes for the spring are offset such that the central axis of torsional spring coincides with the axis of the hinge.  As discussed in the main text and in relation to Fig.~4, we put {a single torsional} spring in a unique fold to make a negative surplus vertex tristable, or we put  {a single torsional} spring in a non-unique fold to make a positive surplus vertex tristable.

\subsection{Measurements of Elastic Energy}

To measure the energy curves for these vertices as shown in Fig.~4, we use an Instron MT-1 torsion tester with a 2.25 N$\cdot$m load cell.  This machine allows us to measure angular displacements with a resolution of 5$\times$$10^{-5}$ rad and torques with a resolution of 0.01 N$\cdot$m.  We have designed special grips to hold the vertex firmly on two adjacent plates while the torsion tester opens/closes the fold between them.

Converting the raw (torque) measurements from the Instron to energy curves such as those in Fig.~4 of main text is a multi-step process.  First, we must determine the rest angle and spring constant of the torsional spring.  We do this by gripping the vertex around the fold with the spring and performing torque vs.~displacement tests opening/closing this fold.  Next, we ensure that the vertex is on the bistable branch and attach it to the instron in a different orientation---this time around the fold opposite the spring.  By performing torque vs.~displacement tests opening/closing this fold, we probe the bistable branch.  Finally, we `pop' the vertex through to the other (monostable) branch and perform opening/closing tests on this fold to probe the monostable branch.

For all of these measurements, we must account for gravity and friction.  Gravity applies a changing torque throughout the experiments on account of the evolving distribution of mass around the folding axis as the vertex opens/closes.  Frictional forces arise because our hinges are imperfect, but are easy to identify/handle as they are constant, rate independent, and switch signs during opening {and} closing.  To account for friction, we do cyclic experiments where we first increase the fold angle $\rho_i$ to its maximum value, then decrease it to its minimum value.  Since the frictional forces are always oriented opposite to the direction of motion, we {average} these forward/backward measurements to obtain curves with friction removed.  To remove gravity, we do two separate experiments:  one with the torsional spring attached to the vertex, and one with the spring {removed}.  By subtracting these two signals, we suppress the contribution from gravity.

After obtaining the friction- and gravity-corrected torque measurements, we simply integrate these to obtain energy curves for the different $\rho_i$ as presented in the main text.  We remark that the precise `zero' position of the Instron has an effect on the asymmetry of the energy curves---if the Instron zero point does not coincide to with the `true' zero point of the fold being probed, the left/right energy curves are offset during integration.  We speculate this is the cause for the asymmetry in our energy curves in Fig.~4e,f, but leave them as is rather than (arbitrarily) picking a new gauge to make them symmetric.

\end{document}